\documentstyle[12pt]{article}

\def\beq{\begin{equation}}
\def\eeq{\end{equation}}
\def\bea{\begin{eqnarray}}
\def\eea{\end{eqnarray}}
\def\ord{\phi (\vec{r},t)}
\def\stat{\phi^s (\vec{r})}
\def\stp{\phi^s (x)}
\def\s0{\phi_0}
\def\p{\partial}
\def\nvec{\vec{\nabla}}
\def\n2{\nabla^2}
\def\la{\langle}
\def\ra{\rangle}
\def\ls{L(t) \sim t^{\frac{1}{3}}}
\def\sd{L(t) \sim t^{\frac{1}{4}}}
\def\st{S(k,t)}
\def\mom{\la k \ra}
\begin{document}
\setcounter{page}{1}
\baselineskip=25pt \parskip=0pt plus2pt
\begin{center}
\begin{Large}
Phase Separation Kinetics in a Model with Order-Parameter
Dependent Mobility
\end{Large}

\vskip0.5cm

by \\
Sanjay Puri$^{1,2,3}$, Alan J.Bray$^1$ and Joel L.Lebowitz$^4$
\end{center}
1. Department of Theoretical Physics, The University, \\
\ \ \ Manchester M13 9PL, U.K. \\
2. Isaac Newton Institute of Mathematical Sciences, Cambridge University, \\
\ \ \ Cambridge CB3 0EH, U.K. \\
3. School of Physical Sciences, Jawaharlal Nehru University, New
Delhi -- 110067, INDIA. \\
4. Departments of Mathematics and Physics, Rutgers University,
New Jersey 08903, U.S.A. \\
\vskip0.5cm
\begin{abstract}
We present extensive results from 2-dimensional
simulations of phase separation kinetics in a model with
order-parameter dependent mobility. We find that the time-dependent
structure factor exhibits dynamical scaling and the scaling function is
numerically indistinguishable from that for the Cahn-Hilliard
(CH) equation, even in
the limit where surface diffusion is the mechanism for domain
growth. This supports the view that the scaling form of the
structure factor is "universal" and
leads us to question the conventional wisdom that
an accurate representation of the scaled structure
factor for the CH equation can only be obtained from a theory
which correctly models bulk diffusion.
\end{abstract}

\pagebreak

\section{Introduction}

When a two-phase mixture in a homogeneous phase is quenched
below the critical coexistence temperature, it becomes
thermodynamically unstable and evolves towards a new equilibrium
state, consisting of regions which are rich in
one or the other constituent of the mixture. The dynamics of this
evolution is referred to as "phase ordering dynamics" and constitutes
a well-studied problem in nonequilibrium statistical
mechanics \cite{jdg}. As a result of these
investigations, there is now a good understanding of 
many aspects of phase ordering
in pure and isotropic binary mixtures. Thus, it is generally accepted that
the coarsening domains are characterised by a unique, time-dependent
length scale $L(t)$, where $t$ is time. Furthermore, 
the nature of the phase ordering process depends critically
on whether or not the order
parameter is conserved. For systems characterised by a
nonconserved order parameter, e.g., ordering of a ferromagnet,
growing domains obey the Lifshitz-Cahn-Allen (LCA) growth law
$L(t) \sim t^{\frac{1}{2}}$ \cite{jdg}. For
systems with a conserved order parameter but no hydrodynamic
effects, e.g., segregation of a binary alloy, the characteristic
domain size obeys the Lifshitz-Slyozov
(LS) growth law $\ls$ \cite{jdg}. 
For systems with a conserved order parameter and
hydrodynamic effects, e.g., segregation of a binary liquid,
there appear to be various regimes of domain growth, depending on the
dimensionality and system parameters \cite{sig, kog}.

As far as the analytic situation is concerned, there is a reasonable
understanding of the nonconserved case for pure and isotropic
systems. In particular, the LCA diffusive growth law has been
derived in some exact models \cite{exa}. In addition,
Ohta et al. and Oono and Puri \cite{ojk} have proposed an analytic form
for the time-dependent structure factor which is in good agreement with
numerical results, though the quality of this agreement
has recently been questioned by Blundell et al. [5]. For the conserved
case, the situation is less satisfactory. There is some
understanding of the growth exponents and one has a good
empirical form for the scaled structure factor -- at least without
hydrodynamics \cite{fra}. However, this functional form is
analytically derivable only in the limiting case where one of the
components is present in a small fraction \cite{jdg}. An outstanding 
theoretical problem in this field
is the calculation of the scaled structure factor for the
conserved case when the two components of the mixture
are present in an equal proportion,
viz., the so-called critical quench \cite{oht}. Our results in this paper
provide some interesting insights on this problem, as we will discuss
later. 

In this paper, we study a model for phase separation dynamics
in systems where the mobility is order-parameter dependent. We will
present detailed numerical results from a simulation of this model.
This paper is organised as follows. In Section 2, we briefly discuss
our model and its static solution. In Section 3, we present
numerical results obtained from our model. Section 4 ends this paper
with a summary and discussion.

\section{Model for Phase-Separating Systems with Order-Parameter
Dependent Mobility}

The dynamics of phase separation is usually described by the 
phenomenological equation
\beq
\frac{\p \ord}{\p t} = \vec{\nabla} \cdot \bigg [ M(\phi) \vec{\nabla}
\bigg ( \frac{\delta H[\ord]}{\delta \ord} \bigg ) \bigg ] ,
\eeq
where $\ord$ is the order parameter at point $\vec{r}$ and time $t$ and is a
measure of the local difference in densities of the two segregating
species, say A and B. In (1), $M (\phi)$ corresponds to the mobility,
which is dependent on the order parameter, in general. The free-energy
functional is usually chosen to be of the standard $\phi^4$-form, viz.,
\beq
H [\ord] = \int \mbox{d} \vec{r} \bigg [ -\frac{1}{2} \ord ^2 +
\frac{1}{4} \ord ^4 + \frac{1}{2} (\vec{\nabla} \ord)^2 \bigg ] ,
\eeq
where we assume that all variables have been rescaled into dimensionless
units; and the system is below the critical temperature. 
The dynamics of Eqs. (1)-(2) drives the order parameter to the
local fixed point values $\phi_0 = \pm 1$, corresponding to (say) A- and
B-rich phases, respectively. The temporal evolution
described by Eq. (1) also satisfies the
conservation constraint that $\int \mbox{d} \vec{r} \ord$ is 
constant in time. 

There have been many studies of Eq. (1) in the limiting case of the
Cahn-Hilliard (CH) equation \cite{ch}, where the mobility is constant,
viz., $M(\phi) = 1$ (in dimensionless units). Numerical studies
of the CH equation and equivalent Cell Dynamical System (CDS) models
\cite{op} demonstrate that late-stage domain growth obeys the
LS growth law we have quoted earlier (i.e., $\ls$). These studies also
clarify the functional form of the scaled structure factor which
characterises the morphology of the coarsening domains.

For deep quenches, it has been pointed 
out by Langer et al. and Kitahara and Imada \cite{lan}
that a more realistic model for phase separation should explicitly 
incorporate an order-parameter dependent mobility of the form
\beq
M(\phi) = 1 - \alpha \phi^2 ,
\eeq
where $\alpha$ parametrises the depth of the quench. At the physical
level, this form of the mobility can be understood as follows. Deep
quenches result in enhanced segregation in that A-rich (or B-rich) domains
are purer in A (or B) than in the case of shallow quenches. Thus, if one
presumes that phase separation occurs by exchanges of neighbouring A- and
B-atoms, the probability of such an exchange in the bulk is drastically 
reduced for deep quenches. This can be mimicked by the order-parameter
dependent mobility in (3) with $\alpha \rightarrow 1$. At the mathematical
level, Kitahara and Imada \cite{lan} have shown
that an order-parameter dependent mobility arises naturally if one
attempts to obtain a coarse-grained model for phase separation from a
master equation description of an appropriate microscopic model, viz., the
Ising model with Kawasaki spin-exchange kinetics \cite{bin}.

The physical effect of the order-parameter dependent mobility is that,
as $\alpha \rightarrow 1$ (which happens for temperature $T
\rightarrow 0$), bulk diffusion is
substantially suppressed because the mobility $M(\phi_0) \rightarrow
0$. Therefore, the effects of surface diffusion are
relatively enhanced. The surface-diffusion mechanism for domain growth
has an associated growth law $\sd$ \cite{fur}, in contrast to the 
evaporation-condensation mechanism which drives asymptotic growth in
the CH equation and gives rise to the LS growth law. Therefore, as 
$T \rightarrow 0$, one expects an extended regime of $t^{\frac{1}{4}}$
growth in the dynamics of Eqs. (1)-(3). This model has been studied 
numerically by various authors \cite{lac} and we will remark on their 
results shortly. Furthermore, Bray and Emmott \cite{be} have analytically
studied phase separation in models with order-parameter dependent mobility
in the limit where one of the components is present in a vanishingly small
fraction. In passing, we should also point out that an order-parameter 
dependent mobility
as in (3) has proven to be a useful way of incorporating the effects
of external fields which vary linearly with distance, e.g., gravity.
However, we will not go into this here and merely refer the interested
reader to Reference \cite{koj}.

In recent work, there was proposed a novel dynamical equation for phase
separation in binary mixtures -- using the master equation
formulation for an Ising model
with Kawasaki spin-exchange kinetics \cite{ppd}. This equation was first
obtained in the context of phase separation in a gravitational field
but does not reduce to the CH equation in the absence of gravity. As
a matter of fact, it takes a form similar to that of Eq. (1), i.e.,
\beq
\frac{\p \ord}{\p t} = \nvec \cdot \bigg [ \big ( 1 - \ord^2 \big ) \nvec 
\bigg ( \frac{\delta H[\ord]}{\delta \ord} \bigg ) \bigg ] ,
\eeq
with the free energy
\bea
H[\ord] & = & \frac{T}{T_c - T} \int \mbox{d} \vec{r}
\frac{1}{2} \bigg [ (1+\ord) \ln (1 + \ord) \nonumber \\
&   & \ \ \ \ \ \ \ + (1-\ord) \ln (1-\ord)
- \frac{T_c}{T} \ord^2 \nonumber \\
&   & \ \ \ \ \ \ \ + \frac{T_c - T}{T} (\nvec \ord)^2 \bigg ] .
\eea
Eqs. (4)-(5) have been cast in a dimensionless form by a rescaling of the
space and time variables analogous to that for the CH equation
\cite{ppd}. (Clearly, this rescaling is not appropriate in
the vicinity of the critical temperature $T_c$.) It is difficult to put
Eqs. (4)-(5) in a parameter-free form because of the additional term in
comparison to the CH equation and the nature of the static solution,
which we discuss below.
The first two terms under the integral sign in (5) are recognised as
the entropy of a noninteracting binary mixture and the next two terms
correspond to the interaction part \cite{zia}.

Eqs. (4)-(5) have the pleasant feature that they explicitly contain the
mean-field static solution $\stat$, which is the solution of
\beq
\stat = \tanh \bigg [ \frac{T_c}{T} \stat + (\frac{T_c}{T} - 1)
\n2 \stat \bigg ] ,
\eeq
where it should be kept in mind that the space variable has been
rescaled. However, we do not
expect our model to be in a different dynamical universality class
from Eqs. (1)-(3). In our model, as $T \rightarrow 0$, the saturation
value of the
order parameter $\phi_0 \rightarrow \pm 1$. This reduces the bulk
diffusion because of the order-parameter dependent mobility
and enhances the time-regime in which one observes
surface-diffusion mediated growth. In the case where surface diffusion
is predominant, we follow the terminology established by Hohenberg
and Halperin \cite{hoh} and refer to our model as "Model S", where S refers
to surface diffusion. In the classification of Hohenberg and Halperin,
the CH equation is referred to as Model B. For shallow quenches, the
saturation value of the order parameter $\phi_0$ is considerably less than
1 and the mobility $M(\phi) ( = 1 - \phi^2)$ is not significantly reduced
in the bulk. In this limit, the dynamics of our model is in the same
dynamical universality class as Model B or the CH equation.

In this paper, we present detailed numerical results from a
simulation of (4)-(5). The purpose of this paper is two-fold. Firstly,
our numerical results improve substantially 
upon existent results \cite{lac} for models with
order-parameter dependent mobility. Secondly, we believe that our
results may be of some relevance to an outstanding theoretical
problem of phase separation dynamics, viz., the computation of the
scaling form of the time-dependent structure factor.

Before we present numerical results, we would like to briefly discuss
the interfacial profile in our model. For this, we need the solution
of the 1-dimensional version of (6), viz.,
\beq
\frac{d^2 \stp}{d x^2} = -\frac{T_c}{T_c - T} \stp + \frac{T}{T_c -T}
\tanh^{-1} (\stp) .
\eeq
Multiplying both sides by $2 (d \stp / dx)$, we can trivially integrate this
equation to get
\bea
\frac{d \stp}{dx} & = & \bigg [ \frac{2T}{T_c - T} \stp \tanh^{-1}
(\stp) + \frac{T}{T_c - T} \ln \bigg ( \frac{1-\stp^2}{1-\s0^2} \bigg ) -
\nonumber \\
&   & \frac{T_c}{T_c - T} (\stp^2 + \s0^2) \bigg ]^{\frac{1}{2}} ,
\eea
where we focus on the profile which goes from $-\s0$ at $x=-\infty$
to $\s0$ at $x=\infty$. A second
integration is only possible numerically and we show the resultant
profiles for $x > 0$ in Figure 1(a) for four different values of $T/T_c$.
This solution has the form $\stp = \s0 f(x/\xi)$ , where $f(y)$ 
is a sigmoidal function
and $\xi$ measures the correlation length or interface thickness
in dimensionless units. An
estimate of $\xi$ is obtained as the distance over which $f(x/\xi)$ rises from
0 to (say) $1/\sqrt{2}$ of its maximum value. The profiles as a function of
the scaled distance $x/\xi$ are shown in Figure 1(b). They do not exhibit a
universal collapse because of a weak dependence of $f(y)$ on the
parameter $T/T_c$. In any case, our interest
in the correlation length is primarily from a numerical standpoint 
in that the
discretisation mesh size in space should not exceed the interface thickness,
which is approximately $2 \xi$.

\section{Numerical results}

We have conducted extensive 2-dimensional 
numerical simulations of (4)-(5) for the
parameter values $T/T_c = 0.2, 0.4, 0.5$ and 0.8, corresponding 
to $\s0 \simeq 0.9999, 0.9857, 0.9575$ and 0.7105,
respectively. We implement a simple Euler discretisation of (4)-(5) on a
lattice of size $N \times N$. The Laplacian and divergence operators in
(4)-(5) are replaced by their isotropically discretised equivalents, involving
both nearest and next-nearest neighbours. The discrete implementation of
our model with order-parameter dependent mobility has the unpleasant
feature that it is unstable for $\phi > 1$ and numerical fluctuations
which cause $\phi$ to become larger than 1 give rise to unphysical
divergences. (This property is common to all such models \cite{lac}.)
For $T/T_c = 0.2 (\s0 \simeq 0.9999)$, this causes a numerical problem
because of the proximity of the saturation value to $\pm 1$.
We circumvent this problem by using a very 
fine mesh size ($\Delta t = 0.001$ and $\Delta x = 0.5$) and by
setting the value of $\phi$ equal to $\s0$ (or $-\s0$) 
whenever it exceeds $\s0$ (or becomes less than $-\s0$). We have confirmed 
that this procedure does not cause
any appreciable violation of order parameter conservation for the
extremely fine mesh we have used. For the higher values of $T$ studied
here, we use the coarser mesh sizes $\Delta t = 0.01$ and 
$\Delta x = 1.0$ and this suffices for our purposes.

Periodic boundary conditions are applied in both directions of our
lattice. For all simulations described here, the initial condition
for the order parameter
consists of a uniformly distributed random fluctuation of amplitude
0.025 about a zero background. This mimics a critical quench from
high temperatures, at which the system is homogeneous but has small
thermal fluctuations.

Apart from evolution pictures and profiles, the statistical quantity
of experimental interest is the time-dependent structure factor
\beq
S(\vec{k},t) = \la \phi (\vec{k},t) \phi (\vec{k},t)^* \ra ,
\eeq
which is the Fourier transform at wave-vector $\vec{k}$ of the order
parameter correlation function. In (9), $\phi (\vec{k},t)$ is the Fourier
transform of $\ord$ and the angular brackets refer to an averaging over
an ensemble of initial conditions. In our discrete simulations, the
wave-vector $\vec{k}$ takes the discrete values $\frac{2 \pi}{N
\Delta x} (n_x,n_y)$, where $n_x$ and $n_y$
range from $-N/2$ to $(N/2) - 1$. 
We present here structure factor data obtained on
$512 \times 512$ systems as an average over 60 independent initial
conditions. The order parameter profiles are hardened before computing
the structure factor, viz., the values of $\phi > 0$ are set equal to 1 and
$\phi < 0$ are set equal to -1. The structure factor is normalised as
$\sum_{\vec{k}} S(\vec{k},t)/N^2 = 1$.
All results presented below are for the spherically averaged
structure factor $S(k,t)$.

Experimentalists are typically interested in whether or not the
structure factor exhibits dynamical scaling \cite{bs}, viz., whether or
not the time-dependence of the spherically averaged structure factor
has the simple scaling form
\beq
S(k,t) = L(t)^d F(kL(t)) ,
\eeq
where $d$ is the dimensionality and $F(x)$ is a time-independent master
function. The interpretation of dynamical scaling is that the
coarsening pattern maintains its morphology but the characteristic length
scale $L(t)$ increases with time. There are many equivalent definitions
(upto prefactors) of the characteristic length scale. We use what is
perhaps the most commonly-used definition, viz., the inverse of the
first moment of the spherically averaged structure factor $S(k,t)$.
Thus, we have $L(t) = \mom ^{-1}$, where
\beq
\mom = \frac{\int_0^{k_m} \mbox{d} k k \st}{\int_0^{k_m} \mbox{d} k
\st} .
\eeq
In (11), we take the upper cut-off $k_m$ as half the magnitude of the
largest wavevector in the Brillouin zone. At these large values of
the wavevector, the 
structure factor has decayed to approximately zero
and the value of $\mom$ is unchanged even if we increase the
cutoff. Of course, one 
could also define a length scale using higher moments of the
structure factor or zeroes of the correlation function. However, in
the dynamical scaling regime \cite{bs}, these definitions are all equivalent.

Figure 2 shows evolution pictures from a disordered initial condition
for the parameter value $T/T_c = 0.2$ (or $\phi_0 \simeq 0.9999$)
and a lattice size $256 \times 256$.
This low value of temperature corresponds to a situation in
which there is almost no bulk diffusion once the order parameter
saturates out to its equilibrium values. In this case,
domain growth occurs via surface diffusion and has an associated
growth law $\sd$ \cite{fur}. Notice that the domain
morphology in this case is considerably different from the morphology
in the usual CH case with the bicontinuous domains being more serpentine and
intertwined in the present case. Figure 3 shows the corresponding
evolution pictures from a $256 \times 256$ lattice
for $T/T_c = 0.5$ (or $\s0 \simeq 0.9575$). These pictures 
are more reminiscent of the CH
morphology. Figure 4 shows the variation of order parameter along a
horizontal cross-section at the middle of the lattice for the
evolution pictures of Figure 2. Figure 5 shows the order parameter
profiles corresponding to the evolution depicted in Figure 3.
These profiles provide a qualitative measure of the thinning
out of defects (viz., interfaces) as the coarsening proceeds.

In Figure 6(a), we superpose data from different times for the 
scaled structure factor $\st \mom^2$ vs. $k/\mom$.
The parameter value is $T/T_c = 0.2$, corresponding to growth mediated
by surface diffusion (i.e., Model S). The structure factor data collapses
neatly onto a master curve, exhibiting the validity of dynamical scaling
in this system. The solid line refers to the scaled structure factor for
the CH equation obtained with the same system sizes and statistics as
described previously. On the scale of this figure, the scaled structure
factor for Model S is coincident with that for the CH
equation except for the first two points after $k=0$, which exhibit violation
of scaling because of finite-size effects. A similar observation has also been
made for the real-space correlation function by Lacasta et al. \cite{lac}.
However, we should stress that the structure factor is a more sensitive
characteristic of phase ordering dynamics than the correlation function. 
Furthermore, our present data
(obtained on $512 \times 512$ systems with 60 independent runs and $\Delta t
= 0.001, \Delta x = 0.5$)
constitutes a considerable improvement over that of Lacasta et al.
\cite{lac}, who used a $120 \times 120$ system with 10 independent 
runs and $\Delta t = 0.025, \Delta x = 1.0$.

Before we proceed, two further
remarks are in order. Firstly, it is interesting
that the structure factors for Model S and the CH model
are numerically indistinguishable, even though the morphologies
are different and domain
growth is characterised by different power laws. Clearly, the
time-dependent structure factor (which is the Fourier transform of the
equal-time correlation function)
is not a sufficiently good measure of the
morphology to discriminate between these two situations and perhaps one
needs to invoke other tools like two-time correlation functions or
higher-order structure factors \cite{blu}. Nevertheless, the
structure factor is an experimentally relevant quantity and the
computation of its analytic form for the
CH equation has been an outstanding problem to date.
Furthermore, it has been believed that a "correct" theory for the scaling
form of the structure factor must properly account for the bulk diffusion
and the LS growth law \cite{oht, maz}. 
However, our numerical results demonstrate that the
scaling form of the structure factor for the conserved case is 
considerably robust and is not affected by the growth exponent or
the underlying growth mechanism, at least for the model we have studied.

The second remark we wish to make concerns the dashed line in Figure 6(a),
which is obtained from a naive application of the theory of
Mazenko \cite{maz}, who developed a Gaussian closure for the CH equation.
The naive Mazenko theory predicts that the asymptotic growth law is
$\sd$ rather than the numerically observed LS law, viz., $\ls$. Because of
the lower growth exponent, it is presumed that the naive Mazenko theory
describes the surface-diffusion growth regime of the CH equation. In the
light of our present results, it is clear that the form of the
scaled structure factor
is largely independent of the mechanism of domain growth.
Unfortunately, as is clear from Figure 6(a), the analytic form 
obtained from the naive Mazenko theory is not correct
in most respects and only gets right the approximate width of the scaling
function. We are presently investigating a Gaussian closure of (4)
to see whether it gives better results for the scaling function.

Figure 6(b) plots the data of Figure 6(a) on a log-log scale and reconfirms
the coincidence of the CH and Model S scaling functions, including the Porod
tail $\st \sim k^{-3}$ for large $k$. At small values of $k$, the 
scaled structure factor for Model S
exhibits a $k^4$-behaviour as in the CH case \cite{yeu}, except for the first
couple of values of $k$, which are probably affected by finite-size
effects. Again, the dashed line is from the naive Mazenko theory and has the
wrong behaviour for small values of $k$, viz., $\st \sim k^{2}$
rather than $\st \sim k^4$. The analytic form 
matches the numerical results in the Porod tail but this may be entirely
fortuitous. Figure 6(c) plots the data of Figure 6(a) on a Porod plot,
viz., $k^4 \st /\mom^2$ vs. $k/\mom$, which highlights features of the
Porod tail. In this case,
our data is not reliable for $k/\mom \geq 2.5$. 
However, upto that point, the scaled
form factors for the Model S and CH cases are again
indistinguishable, including the first valley after the peak
\cite{oht}. 

Similar results for the scaled structure factor are found for higher
values of temperature $T$ also. This is not surprising as the morphology 
for our model goes over to that for the CH equation at higher values of the
temperature (see Figure 3). For the sake of brevity, we do not show 
structure factor data for higher values of $T$.

Figure 7(a) shows the time-dependent length scale $L(t)$ as a function of
dimensionless time $t$ for four different values of temperature
($T/T_c = 0.2, 0.4, 0.5$ and 0.8) in our model. Recall that surface
diffusion effects are enhanced as $T$ is lowered because $\s0 
\rightarrow 1$ as $T \rightarrow 0$. For purposes of
comparison, we have also included the length scale data for the CH
equation. Figure 7(b) is a log-log plot of the data in Figure 7(a). We
use a fitting routine to fit a straight line to the data. The
resultant exponents (denoted as $x$) for the CH equation and the case with
$T/T_c = 0.8$ are identical,
viz., $x=0.33$. On the other hand, for $T/T_c = 0.2$, we again get a straight
line but the associated growth exponent is 0.25, which is associated
with domain growth via surface diffusion \cite{fur, lac}. For intermediate
values of $T/T_c$ (viz., 0.4 and 0.5), we do not get a good linear fit as
the length scale is in a transition regime between $\sd$ and $\ls$.

\section{Summary and Discussion}

        Let us end this paper with a brief summary and discussion of
our results. We have presented detailed results from an extensive
numerical simulation of a model with order-parameter dependent
mobility. We expect this model to be in the same dynamical
universality class as other models with order-parameter dependent
mobility \cite{lan, lac} but it has the additional pleasant feature that it
explicitly contains the mean-field static solution.

        Because of the large system sizes and extensive averaging
employed by us, we are able to obtain the best numerical results on
such systems to date. The salient features of our results are as
follows. In the parameter regime where surface diffusion drives
domain growth, the morphology of evolving patterns is more
serpentine than that in the CH equation. However, the scaling form of
the time-dependent structure factor for surface-diffusion mediated
growth appears to be numerically identical to that
for the CH equation, including the Porod tail and the small-$k$
behaviour. This numerical result casts doubts on the conventional
wisdom that a "correct" theory for the scaling form of the CH
structure factor must contain the correct growth law and properly
model the bulk diffusion field. As a matter of fact, we are led to
speculate that the scaling form for the conserved case may be
dictated by more
general considerations, e.g., domain-size distributions, etc. This is
an approach we are presently pursuing in an attempt to obtain a better
understanding of the functional form of the structure factor for the
conserved case.

We are also interested in examining other models of phase
separation to see whether they give rise to similar results for
the scaled structure factor. In particular,
Giacomin and Lebowitz \cite{gia} have recently studied an
Ising model on a cubic lattice with Kawasaki spin-exchange
kinetics which satisfies detailed balance. The spins interact
via a long-ranged Kac interaction potential of the form
$V(r_{ij}) = \gamma^d J(\gamma r_{ij})$, where $r_{ij}$ is the
distance between spins $i$ and $j$; $\gamma$ is a parameter; and
$d$ is the dimensionality. In the limit $\gamma \rightarrow 0$,
Giacomin and Lebowitz rigorously obtain an exact nonlinear
evolution equation for phase separation. Their model is of the
same form as Eqs. (4)-(5) but contains a nonlocal interaction term,
instead of the gradient square term in (5). They argue that
this exact equation gives results for interface motion which are
similar to those obtained from the CH equation. We are
interested in examining whether or not this exact equation is in
the same dynamical universality class as the CH equation.

Finally, we should point out that the difference in
morphologies between Model S and the CH equation must show up at some
level, e.g., two-time correlation functions or
higher-order structure factors \cite{blu}. This is another
question we are presently interested in. Nevertheless, this possible
difference in two-time correlation functions or
higher-order structure factors does not detract 
from the relevance of the fact that the scaled form of
the conventional structure factor is very robust. After all, the
conventional structure factor is the primary quantity of experimental,
numerical and theoretical interest. 

\section*{Acknowledegments}        

SP is grateful to Alan Bray for inviting him to Manchester, where most
of the numerical calculations described in the text were completed.
He is also grateful to the Newton Institute, Cambridge, for its
generous hospitality during a period over which this work was
completed. Finally, he would like to thank A.-H.Machado, C.Yeung
and R.K.P.Zia for useful discusions and A.-H.Machado for sending him
copies of relevant papers. JLL and SP thank G.Giacomin for
useful discussions. JLL was supported by NSF Grant NSF-DMR 92-134244-20946.

\newpage
\begin{center}
{\bf Figure Captions}
\end{center}

\begin{itemize}
\item[Figure 1 :] (a) Static wall solutions of the model described in the
text (Eqs. (4)-(5)). The solutions are obtained by numerically solving
(8). We plot the profile $\stp/\s0$ vs. $x$ for $x > 0$ (where $\s0$ 
is the saturation value)
for four values of the temperature $T$, viz., $T/T_c = 0.2, 0.4,0.5$ and
0.8. \\
(b) Same as (a) except the distance $x$ is scaled by a
correlation length $\xi$, which is defined as the distance over which
the wall profile rises to $1/\sqrt{2}$ of its maximum value.

\item[Figure 2 :] Evolution pictures from a disordered initial
condition for an Euler-discretised version of (4)-(5) on a $256 \times 256$
latice. Regions with positive order parameter are marked in black and
those with negative order parameter are not marked.
The parameter value is $T/T_c = 0.2$, corresponding to a situation
in which surface diffusion is the primary mechanism of domain
growth. The discretisation mesh sizes are $\Delta t = 0.001$ 
and $\Delta x = 0.5$. Periodic
boundary conditions are applied in both directions. The initial
condition consists of uniformly-distributed random fluctuations of
amplitude 0.025 about a zero background, corresponding to a critical
quench. The evolution pictures are shown for dimensionless times 1000, 2000,
4000 and 10000.

\item[Figure 3 :] Similar to Figure 2 but for the parameter value 
$T/T_c = 0.5$.

\item[Figure 4 :] Order parameter profiles for the evolution
depicted in Figure 2. The profiles are measured along a horizontal
cross-section at the centre of the vertical axis.

\item[Figure 5 :] Order parameter profiles for the evolution
depicted in Figure 3. The cross-section is the same as that for Figure
4.

\item[Figure 6 :] (a) Superposition of scaled structure factor
data from a simulation of (4)-(5) with $T/T_c = 0.2$, corresponding to the
surface-diffusion case. We plot $\st \mom^2$ vs. $k/\mom$ for data
from dimensionless times 2000, 3000, 4000 and 10000. The structure
factor is computed on a $512 \times 512$ lattice as an average over 60
independent initial conditions. It is normalised as described in the
text and then spherically averaged. The first moment of $\st$ is
denoted as $\mom$ and measures the inverse of the characteristic
length scale. The solid line is a scaled plot of structure 
factor data from the CH
equation at dimensionless time 10000. Finally, the dashed line is an
analytic form obtained from a naive application of
Mazenko theory \cite{maz}, which yields
the domain growth law $\sd$. \\
(b) Plot of data from (a) on a log-log scale. The Porod tail
is extracted by hardening the order parameter field before computing
the structure factor. \\
(c) Porod plot (viz., $k^4 \st /\mom^2$ vs. $k/\mom$) 
for the data from (a). This plot
highlights the features of the Porod tail. Unfortunately, our data in
this plot exhibits large fluctuations for $k/\mom \geq 2.5$.

\item[Figure 7 :] (a) Characteristic domain size $L(t)$ plotted as a
function of dimensionless time for our model in (4)-(5) 
with $T/T_c = 0.2, 0.4,
0.5$ and 0.8. For comparison, we also present length scale data from a 
simulation of the CH equation. The length scale is obtained as the
inverse of the first moment of the structure factor $\mom$. \\
(b) Data from (a), plotted on a log-log scale. We use a
fitting routine to fit a linear function to the length scale data.
The resultant fit (wherever reasonable) is shown on the appropriate
data set as a solid line and the corresponding exponent (denoted as
$x$) is specified on the figure.        
\end{itemize}

\end{document}